\documentclass[fleqn,usenatbib]{mnras}

\usepackage[T1]{fontenc}

\DeclareRobustCommand{\VAN}[3]{#2}
\let\VANthebibliography\thebibliography
\def\thebibliography{\DeclareRobustCommand{\VAN}[3]{##3}\VANthebibliography}

\usepackage{graphicx}	
\usepackage{amsmath}	
\usepackage{amssymb}	
\usepackage{ulem}

\title[Fourier method for focusing]{A Fourier method for the
  determination of focus for telescopes with stars}

\author[C. Y. Tan and B. Schulz]{
C.Y. Tan,$^{1}$\thanks{E-mail: cytan299@yahoo.com}
and B. Schulz$^{2}$\thanks{E-mail: Schulz.Benjamin@googlemail.com}
\\
$^{1}$Aurora, IL60504, USA\\
$^{2}$81249 M\"unchen, Germany 
}

\date{Accepted XXX. Received YYY; in original form ZZZ}

\pubyear{2021}

\begin{document}
\label{firstpage}
\pagerange{\pageref{firstpage}--\pageref{lastpage}}
\maketitle

\begin{abstract}
  We introduce a Fourier method (Fm) for the determination of best
  focus for telescopes with stars. Our method fits a power function, that we will
  derive in this paper, to a set of images taken as a function of
  focuser position. The best focus position is where the power is
  maximum. Fm was first tested with small refractor and
  Schmidt-Cassegrain (SCT) telescopes. After the successful small
  telescope tests, we then tested Fm with a 2~m
  Ritchey-Chr\'etien-Coud\'e (RCC).  Our tests show that Fm is immune
  to the problems inherent in the popular half-flux diameter method.
\end{abstract}




\section{Introduction}\label{sec:intro}

One very important aspect in the collection of good astronomical data
is the quality of the focusing of a telescope. A typical observing session
can last between 1 to 12 hours depending on the season and latitude.
During this time, the ambient temperature will fluctuate. Temperature
changes induce strain in optical components and the materials of the
optical train from which a telescope is built and this can affect the
focal point of the telescope. As a result, the position of the camera
sensor connected at the end of the telescope has to be adjusted along
with the temperature changes. To solve this problem, multiple focusing
runs are done throughout the observing session to keep the images
sharp.  Without good and precise focusing, the collected data
can be less than optimum for the given atmospheric conditions.

A quick survey of commercial and free autofocus programs (for example,
focusmax, SGP, SharpCap\footnote{ The full width half max method
  (FWHMm) is used by SharpCap. FWHMm is the predecessor to HFDm.})
will show that the method that is employed by all these focusing
programs is the half-flux diameter method (HFDm) or its
predecessor.\footnote{To the authors' knowledge, there is only one
  program, called SharpLock, that does not use HFD. It uses the
  astigmatism of a star image for determining focus. See
  \cite{Baudat}.}  As the telescope gets into focus, the HFD measured
for one single star or the average HFD from a star field decreases,
and as the telescope gets out of focus, the HFD increases.  Thus, the
minimum of the HFD is the focus position. In principle, although the
calculation of HFD is trivial, there are a few weaknesses with
focusing using HFDm.  We will highlight them throughout this paper and
in section~\ref{sec::HFDm}.

In this paper, we will derive first the model used by the Fourier
method (Fm) and then apply Fm to determine the focus of several
telescopes. We start by applying Fm to small refractor and
Schmidt-Cassegrain (SCT) telescopes. After the successful small
telescope tests, we tested Fm with the 2~m Ritchey-Chr\'etien-Coud\'e
(RCC) telescope installed in the Rozhen National Astronomical
Observatory (Rozhen NAO) in
Bulgaria.\footnote{This is the largest telescope in south-east
  Europe.} In all cases, we will compare the Fm focus results to the
HFDm focus results.

We want to note that we are not the first who use Fourier space for
autofocus purposes, for example, \cite{fourierfocus}. Additionally, we
are also not the first authors who use autofocus methods in astronomy
that do not rely on the detection of individual stars, for example,
see the review by \cite{popowicz}. However, all previous approaches only show
that there is some focus function which attains a special point
(e.g. a global maximum) at focus. Those methods can only select the
sharpest image from a series of images because the precise form of the
focus function for a given image type is not known. In contrast, the
novelty of our paper is that we compute the explicit form of the focus
function for an image type that is very common in astronomy, namely
images which contain stars. We will show that we can use our focus
function to compute the focus position with a curve fit from a few
defocused images. Our real world examples will also show that as long
as there are stars in the image, our focus function is not sensitive
to the presence of additional diffuse structures that may come from
nebul\ae\ or galaxies. Fm is also insensitive to telescope designs
that have obstructions in the light path from a secondary mirror.
Furthermore, we will demonstrate a curve fitting algorithm which
employs methods of robust statistics, that considerably reduces the
influence of any remaining seeing problems and optical errors in the
images. Finally, we will use our algorithm in connection with
different statistical estimators to prove that Fm yields better
results than the commonly used HFDm.

\section{Basic idea of the Fourier method}

\begin{figure*}
  \centering
    \includegraphics*[width=\textwidth]{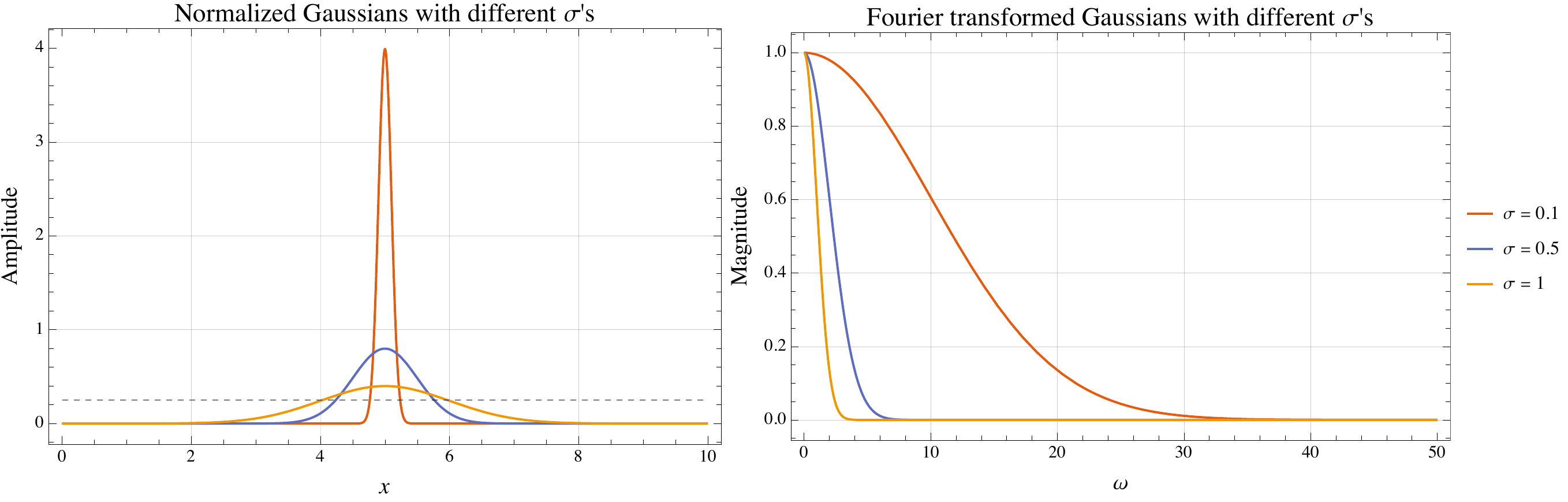}
    \caption{These two plots show the well-known feature that when the
      Gaussian is thin in real space, it is wide in Fourier space. The
      dashed horizontal line in real space space will be used for our
      discussion in section~\ref{sec::saturated}.}
  \label{gaussian.pdf}
\end{figure*}

The basic idea of Fm is to recall the well-known fact that the Fourier
transform of a Gaussian is a Gaussian. And that a thin Gaussian
in real space becomes a wide Gaussian in Fourier space.  For example,
we can demonstrate this with a normalized 1-D Gaussian in real space
\begin{equation}
  g(x,a,\sigma) =  \frac{a}{\sqrt{2\pi}\sigma}{\rm e}^{-x^2/2\sigma^2}
  \label{th.1}
\end{equation}
where $a$ is the amplitude, and $\sigma$ is the standard
deviation. Then its Fourier transform when it is centered at $x_0$ is
\begin{equation}
  {\rm FT}[g(x-x_0,  a,\sigma)] =
  \tilde g(\omega, a, \sigma ) = a{\rm e}^{-\sigma^2\omega^2/2}{\rm e}^{-i\omega x_0}
  \label{th.2}
\end{equation}
where we have used the engineer's definition of the Fourier transform
\begin{equation}
  \tilde f(\omega) = \int_{-\infty}^\infty f(t) {\rm e}^{-i\omega t}\; dt
  \label{th.3}
\end{equation}
We can plot $g$ and $\tilde g$ for $\sigma\in\{0.1,0.5,
1\}$, $a=1$ and $x_0
=5$ to demonstrate the basic idea. From Fig.~\ref{gaussian.pdf},  we
can see that the thinner or sharper the Gaussian is in real space the
wider it is in Fourier space. By integrating under the Fourier space
Gaussian, we can see that the sharper the Gaussian the larger the integral
is in Fourier space. This forms the basis of the Fm method.

We notice that, unlike HFDm that gets smaller as the focus is reached, Fm
gets larger as the focal point is reached. By integrating under the curve in
Fourier space, which is in effect low pass filtering, means that Fm is
less noisy by design.

\section{Gaussian star fields}

We will assume that the exposure times at which the measurements for
focusing are taken are long enough, so that on average, the intensity
profiles of the stars can be described approximately by a Gaussian,
i.e.~we will ignore the transfer function of the telescope or we can
think of this approximation as the consideration of the central peak
of the PSF (point spread function) only.  By making the Gaussian
approximation, we will be able to compute a curve for the intensity
profile of a star field in Fourier space.

Later, in section~\ref{sec::examples}, we will show that $4$ to $8$~s
exposures are sufficient for telescopes of varying focal lengths and
F-numbers to minimize any star shape variations from Gaussian that are
due to atmospherics. Any remaining deviations can be handled by our
outlier removal procedures with robust statistics.
%

Therefore, let us suppose that a star seen on a CCD/CMOS camera at location
$(x_j,y_j)$ can be modeled as
a 2-D Gaussian
\begin{equation}
g(x-x_j,y-y_j; a_j,\sigma_j) =
\frac{a_j^2}{2\pi\sigma_j^2}{\rm e}^{-\frac{(x-x_j)^2+(y-y_j)^2}{2\sigma_j^2}}
\label{gsf.1}                                        
\end{equation}
where we have, for simplicity, assumed that the star is round,
i.e.~its $\sigma$'s are equal in $x$ and $y$. We are also ignoring the
boundaries set by the finite size of the CCD/CMOS chip.  The 2-D
Fourier transform of the above is
\begin{equation}
  \begin{aligned}
&  {\rm FT}[g(x-x_j, y-y_j; a_j,\sigma_j)]   = \tilde g(\omega_x,
                                              \omega_y;
                                              a_j,\sigma_j)\\
  &= \int_{-\infty}^\infty dx\;
  \int_{-\infty}^\infty dy\; g(x-x_j,y-y_j; a_j,\sigma_j) {\rm e}^{-i\omega_x
                                        x}{\rm e}^{-i\omega_y y}\\
                                      &=a_j^2{\rm
                                        e}^{-\frac{1}{2}\sigma_j^2(\omega_x^2+\omega_y^2)}{\rm
                                        e}^{-i(\omega_xx_j+\omega_yy_j)}
\label{gsf.2}                                        
\end{aligned}
\end{equation}

Let us suppose that there are $N$ stars seen by the camera, then by 
the superposition of Fourier transforms, we have
\begin{equation}
\begin{aligned}
  \tilde G(\omega_x, \omega_y) &=\sum_{j=1}^N \tilde g(\omega_x,
                                   \omega_y; a_j,\sigma_j) \\
  &= \sum_{j=1}^N
  a_j^2{\rm e}^{-\frac{1}{2}\sigma_j^2(\omega_x^2+\omega_y^2)}{\rm e}^{-i(\omega_xx_j+\omega_yy_j)}
\end{aligned}
\label{gsf.3}
\end{equation}
We are only interested in the square of the amplitude of the star
field in Fourier space, this means that
\begin{equation}
\begin{aligned}
&  |\tilde G|^2 = \tilde G\tilde G^* \\
  &=
\sum_{j,k=1}^N  a_j^2a_k^2{\rm e}^{-\frac{1}{2}(\sigma_j^2 +\sigma_k^2)
                 (\omega_x^2+\omega_y^2)}{\rm
    e}^{-i\omega_x(x_j-x_k)}{\rm e}^{-i\omega_y(y_j-y_k)}
  \\
  &= \sum_{j=1}^Na_j^4{\rm e}^{-\sigma_j^2(\omega_x^2+\omega_y^2)} 
  \\
&\quad  +  \sum_{j=1}^N\sum_{k\ne j, k=1}^Na_j^2
    a_k^2{\rm e}^{-\frac{1}{2}(\sigma_j^2
    +\sigma_k^2)(\omega_x^2+\omega_y^2)}
  \\
  &\qquad
    \times\cos\left[\omega_x(x_j-x_k)+\omega_y(y_j-y_k)\right]
  \end{aligned}
\label{gsf.10}
\end{equation}
where we have separated out the contribution from the same star and
the contribution from the neighbors of each star. In general, the
double sum does not vanish.

\subsection{Projection onto $\omega_x$-axis}

At this point of our analysis, we are still working in 2-D Fourier
space. We can make it 1-D by projecting the 2-D power distribution into a 1-D
distribution by integrating Eq.~\ref{gsf.10} in $\omega_y$,
\begin{equation}
  \begin{aligned}
    {\cal P}(\omega_x) &= \int_{-\infty}^\infty|\tilde G|^2\; d\omega_y
    =
\sqrt{\pi}\sum_{j=1}^N\frac{a_j^4}{\sigma_j}{\rm e}^{-\sigma_j^2\omega_x^2}\\
&+\sqrt{2\pi}\sum_{j=1}^N\sum_{k\ne j,
  k=1}^N\frac{a_j^2a_k^2}{(\sigma_j^2+\sigma_k^2)^{\frac{1}{2}}}{\rm e}^{-\frac{(y_j-y_k)^2}{2(\sigma_j^2+\sigma_k^2)}}
{\rm e}^{-\frac{1}{2}(\sigma_j^2 + \sigma_k^2)\omega_x^2}\\
&\qquad\times\cos\omega_x(x_j-x_k)
\end{aligned}
\label{pj.1}
\end{equation}
It is obvious from the above that there is no analytic formula for
$\cal P$. However,  we will derive a phenomenological formula
good for fitting Fm in the next section.

\subsection{Phenomenological model}

Since we only want to find the form of the fit formula used in Fm, we
will replace every $a_j$ and $\sigma_j$ by the mean of the amplitude
and sigma of the stars in the image, i.e.~$a_j\rightarrow \bar a$ and
$\sigma_j \rightarrow \bar\sigma$.\footnote{The idea of replacing
  statistical variables by their means to glean insight into the
  behavior of the bulk is not new. This is the idea behind mean field
  theory. Whether it is plausible here is demonstrated by the good
  behavior and fits to empirical data.} This means that Eq.~\ref{pj.1}
becomes
\begin{equation}
  \begin{aligned}
  \bar{\cal P}(\omega_x) &= N\sqrt{\pi}\frac{\bar
    a^4}{\bar\sigma}{\rm e}^{-\bar\sigma^2\omega_x^2} \\
  &+\sqrt{\pi} \frac{\bar
    a^4}{\bar\sigma}{\rm e}^{-\bar\sigma^2\omega_x^2}
\sum_{j=1}^N\sum_{k\ne j,
  k=1}^N  
{\rm e}^{-\frac{(y_j-y_k)^2}{4\bar\sigma^2}}\cos\omega_x(x_j-x_k)
\end{aligned}
\label{ph.1}
\end{equation}
Now, we can integrate in $\bar{\cal P}$ over $\omega_x$ to get the power contained
in the image as a function of the sigma star size in the image
\begin{equation}
  \begin{aligned}
  \bar{\cal P}_{\rm image}(\bar\sigma) &=
  \int_{-\infty}^\infty\bar{\cal P}(\omega_x)\; d\omega_x \\
  &=
\pi\frac{\bar
    a^4}{\bar\sigma^2}\left( N+ 
\sum_{j=1}^N\sum_{k\ne j,
  k=1}^N  
 {\rm e}^{-\frac{(y_j-y_k)^2}{4\bar\sigma^2}}{\rm
   e}^{-\frac{(x_j-x_k)^2}{4\bar\sigma^2}}\right)
\end{aligned}
\label{ph.2}
\end{equation}
For the double sum, we will make the assumption that the distance
between stars is much larger than the diameter of the star,
i.e.~$|x_j-x_k|, |y_j-y_k| \gg 2\bar\sigma$ in the image. We can think
of it as the sparse star field approximation. (We will discuss the
dense star field in section~\ref{sec::dense} and in
Appendix~\ref{app::dense}.)  This means that the double sum becomes
\begin{equation}
  \sum_{j=1}^N\sum_{k\ne j,
  k=1}^N  
{\rm e}^{-\frac{(y_j-y_k)^2}{4\bar\sigma^2}}{\rm e}^{-\frac{(x_j-x_k)^2}{4\bar\sigma^2}}
\approx   0 
\label{ph.3}
\end{equation}
if $(x_j-x_k)^2, (y_j-y_k)^2 \gg 4\bar\sigma^2$ and Eq.~\ref{ph.2} becomes
\begin{equation}
    \bar{\cal P}_{\rm image}(\bar\sigma)\approx \frac{N\pi \bar
      a^4}{\bar\sigma^2}
    \label{ph.4}
  \end{equation}
  
In practice, the size of the star seen on the image will never be zero
at focus but will have some minimum size. Let us define $\bar\sigma_0$
to be the average size of the stars when in-focus. When we take this
observation into account, Eq.~\ref{ph.4} becomes
\begin{equation}
  \bar{\cal P}_{\rm image}(\Delta\sigma) \approx \frac{N\pi \bar
      a^4}{(\bar\sigma_0 + \Delta\sigma)^2}
\label{ph.5}
\end{equation}
This is the phenomenological model that we will adopt as the
power contained in an image for a given mean star size.

\subsection{Fm formula}
\label{sec::Fmformula}

In Fm, we have to model the behavior of $\bar\sigma$ when the focuser
is racked in and out. The functional form between focuser position,
$z$, to average star size is modeled as a hyperbola\footnote{For
  example, in HFDm, the model that connects the HFD to focuser
  position is a hyperbola which has the form $x^2/a^2 - y^2/b^2 =
  1$. We use the hyperbola to fit to the HFD data in
  section~\ref{sec::compareHFD}.}
\begin{equation}
  (\bar\sigma_0 + \Delta\sigma)^2 = K(z-z_0)^2+ \bar\sigma_0^2
  \label{fm.1}
\end{equation}
where $z_0$
is the location of focus and $K$ is a scaling that converts position
to star size. This is clearly not a linear model between star
size and focuser position.

We can substitute the above into Eq.~\ref{ph.5} to get
\begin{equation}
  \bar{\cal P}_{\rm Fm}(z; \alpha, \beta, \gamma, z_0) =
  \frac{  \alpha}{(z-z_0)^2+\gamma} +\beta
  \label{fm.2}
\end{equation}
where $\alpha = N\pi \bar a^4/K$, $\beta$, $\gamma = \bar\sigma_0^2/K$
and $z_0$ are the variables that we need to fit to the measurements of
$\bar{\cal P}_{\rm FM}$ as a function of the focuser position $z$. We
have introduced the necessary offset variable $\beta$ because it is
needed to account for the power in the background noise, which we
assume is white, inherent in all CCD/CMOS images. We can take
advantage of the white noise assumption for calculating $\beta$ by
equating it to the average of the high frequency components of the
power spectrum. We can do this because (a) for bandwidth limited white
noise, the average power is a constant independent of frequency and
(b) in practice, most of the frequency components of the star are in
the low frequency parts of the spectrum. For example, see
Fig.~\ref{gaussian.pdf}.

\subsection{Dense star field}
\label{sec::dense}

\begin{figure}
  \centering
    \includegraphics*[width=\columnwidth]{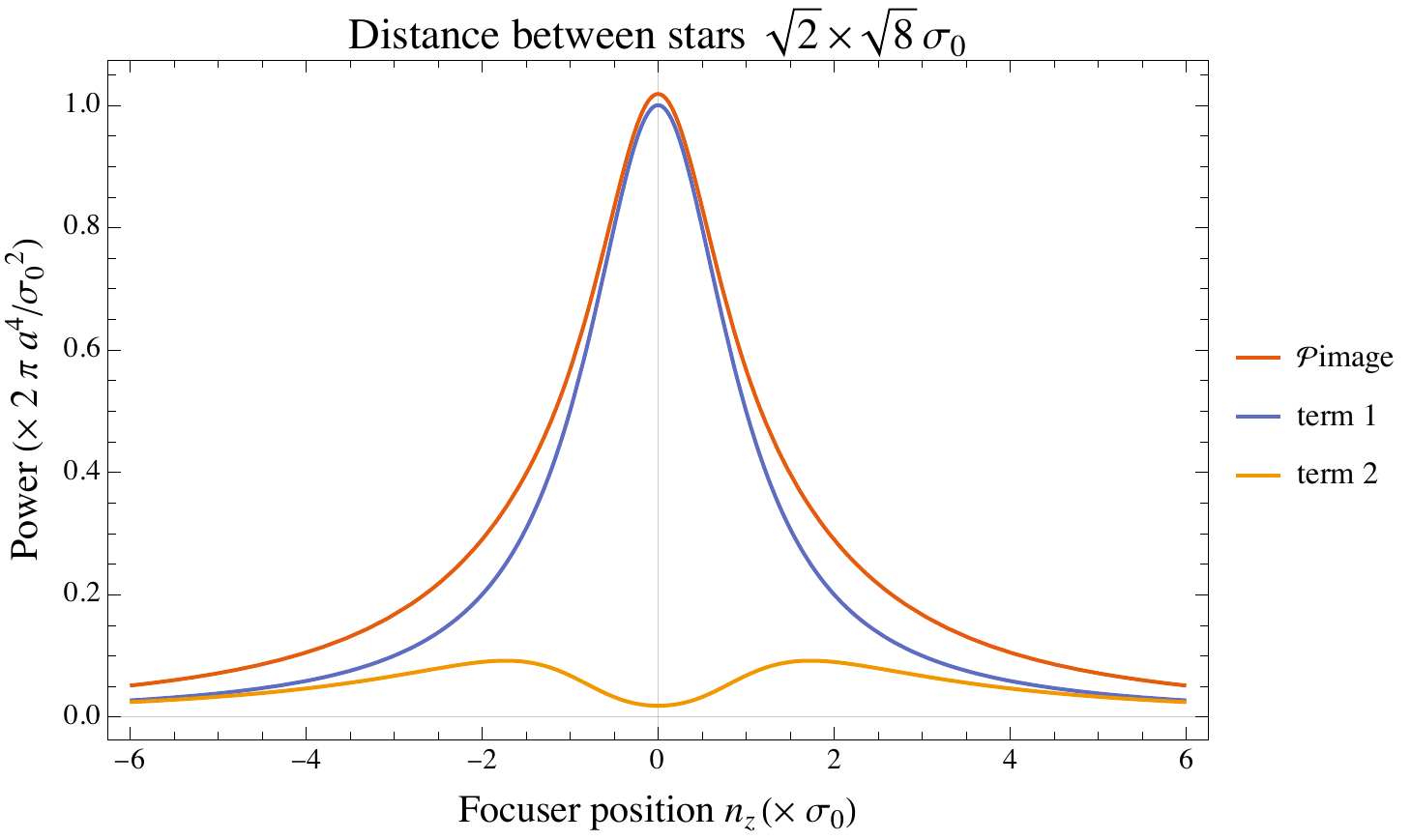}
    \caption{This is an example of the effect of the double sum of
      Eq.~\ref{eq:doublesum} when we have two stars spaced
      $\sqrt{8}\bar\sigma_0$ in both the $x$ and $y$ directions. The
      double sum (term~2) clearly shows the dip in its value at
      focus. However after adding in term~1, the result does not have
      a dip.
}
  \label{g8a.pdf}
\end{figure}

In this paper, although we will be applying the sparse star field
approximation to all of our examples in section~\ref{sec::examples}, we
will digress here and consider the effects of a dense star field.

For a dense star field, the double sum in Eq.~\ref{ph.3} does not
vanish, i.e.~we have, from Eq.~\ref{ph.2}
\begin{alignat}{5}
    \bar{\cal P}_{\rm image}(\bar\sigma^2) &= \frac{N\pi \bar
      a^4}{\bar\sigma^2} &+& \frac{\pi\bar a^4}{\bar\sigma^2}\sum_{j=1}^N\sum_{k\ne j,
  k=1}^N  
 {\rm e}^{-\frac{(y_j-y_k)^2}{4\bar\sigma^2}}{\rm
   e}^{-\frac{(x_j-x_k)^2}{4\bar\sigma^2}}\nonumber\\
 & = \hbox{term 1} &+& \quad\hbox{term 2}\label{eq:doublesum}
\end{alignat}  
The double sum (term 2) has the interesting
property that at focus, its value is smaller than when it is slightly
out of focus. The proof can be read in Appendix~\ref{app::dense}.  We
can illustrate this observation by considering the special case when
$N=2$ and the distance between them projected onto the $x$ and $y$
axes to be $\sqrt{m}\bar\sigma_0$ for $m=8$ and $K=1$. We plot term~1, term~2
and $\bar{\cal P}_{\rm image}(\bar\sigma^2)$ in Fig.~\ref{g8a.pdf} as a
function of focuser position $z=n_z\bar\sigma_0$ for $z_0 =0$. From this
figure, we can clearly see the dip in term~2 at focus.  But after
adding term~1, which is the sparse star approximation (Eq.~\ref{ph.4}),
the dip does not appear in the final sum.

\section{Fitting and outlier removal}
\label{sec::fit}

A common method for curve fitting to data which approximately
fulfills a known differentiable function is the Levenberg-Marquardt (LM)
algorithm (\cite{Levenberg,Marquardt}). This algorithm computes the sum
of the squares of the residuals, i.e.~the sum of the squared
difference of a given function and the data. The algorithm proceeds
as an iterative procedure searching for a local minimum of this
residual. Unfortunately, because the iterative procedure only searches
for a local minimum, the LM algorithm is known to
depend on the precise choice of the initial data.  In our case, we
often have considerable outliers that come from seeing and high
altitude clouds. These outliers make it difficult to start the
LM algorithm with the required initial guess of the
focuser position.

If we insert $(\bar{\sigma}_0+\Delta\sigma)^2$ into
Eq.~\ref{eq:doublesum} and consider
$\bar{\cal{P}}_{\rm image}(\Delta\sigma)\equiv \bar{\cal{P}}_{\rm
  image}((\bar\sigma_0+\Delta\sigma)^2)$ as a function of
$\Delta\sigma$, we observe that both term 1 of Eq. \ref{eq:doublesum}
and $\bar{\cal{P}}_{\rm image}(\Delta\sigma)$ have their maximum when
$\Delta\sigma = 0$. Furthermore, when we compare $\bar{\cal P}_{\rm image}$ of
Eq. \ref{eq:doublesum} to term~1, $\bar{\cal P}_{\rm image}$ has a higher
peak and tails.
Therefore, we can can start with an initial fit with just term~1. For
all of our examples, the fitting of just term 1 turns out to be
sufficient for determining the focus position.

We can convert Eq.~\ref{fm.2} into a line if we consider
\begin{equation}
  Y(\zeta, \alpha, \gamma) =\frac{1}{\bar{\cal P}_{\rm
      Fm}(\zeta;\alpha,\beta,\gamma)-\beta}={\frac{1}{\alpha}\zeta+\frac{\gamma}{\alpha}}
\label{eq::powerfit}
\end{equation}
where we have substituted $\zeta=(z-z_0)^2$ in
$\bar{\cal P}_{\rm Fm}$. This enables us the use of simple linear regression
techniques for estimating the focus position.

In Eq.~\ref{eq::powerfit}, we have subtracted the parameter
$\beta$. In the practical implementation, we get $\beta$ from the
magnitudes of the columns that represent white noise in the Fourier
transformed image data. After $\beta$ is subtracted from the
magnitudes of the Fourier data, the remaining positive magnitudes are
then summed. This also implies that the result of our procedure is
generally different than what can be achieved by a simple summation of
the magnitudes of the image's pixels in real space. The Fourier method
is distinctive in that it allows us to easily and quickly remove most
of the noise. Without the ability to easily subtract the $\beta$ term
from the Fourier data, we would have to use the LM algorithm on
$\bar{\cal P}_{\rm Fm}(\zeta;\alpha,\beta,\gamma)$. However, we have
no information about reasonable initial values for
$\zeta,\alpha,\beta,\gamma$ and thus the outcomes of the LM algorithm
would not be reliable. It is only with the removal of $\beta$ from
$\bar{\cal P}_{\rm Fm}(\zeta;\alpha,\beta,\gamma)$, that we can use
powerful linear regression methods on $Y(\zeta,\alpha,\gamma)$ which
do not depend as much on the chosen initial values for
$\zeta,\alpha,\gamma$ when compared to the LM algorithm. In
section~\ref{sec::shaping}, we will use the LM algorithm only as a
second stage to account for exponential functions in the double sums
of Eq.~\ref{ph.3}. The LM algorithm is then used with initial values
that we have acquired from the linear regression.

We expect that $Y(\zeta,\alpha,\gamma)$ only resembles a parabola
close to the focus position, because there are small corrections from
the double sum in Eq.~\ref{ph.3}. It is therefore not surprising that
we have found, in practice, that linear regression alone is insufficient
for obtaining $z_0$ from a fit of $Y(\zeta,\alpha,\gamma)$ because we
have to remove the outliers in the data. See the examples in
Figures~\ref{gridngc7023.pdf} -- \ref{gridC8.pdf}.

In order to fit our data appropriately and remove the remaining
outliers from our rather small set of measurement points, we will use
a modified version of a RANSAC algorithm (see \cite{ransac} for the
original description of RANSAC) that works deterministically and uses
robust estimators (\cite{Benni}) to solve the least trimmed squares
problem (see \cite{robustregression}).  We will describe this algorithm in
detail below.

Suppose we are given a set $\cal M$ that contains $N_{\cal M}$ data
points with coordinates $\left(z_i,Y_i\right)$ for
$i=1,\ldots, N_{\cal M}$ where $z_i\in\mathbb{Z}$ is the focuser
position and $Y_i$ is from Eq.~\ref{eq::powerfit}.  We then choose to work
with a smaller number of data points, $n\le N_{\cal M}$ from $\cal M$
so that we can add in points that have not been originally selected
using robust estimators. We will call the reduced data set that we will work with
${\cal A}_\ell=\left\{\left(z_{\ell_{k}}
    ,Y_{\ell_{k}}\right),k=1,\ldots, n\right\}$.  For simplicity, let
us assume that the data points in ${\cal A}_\ell$ have been sorted so
that $z_{\ell_1} \le z_{\ell_2}\le \ldots \le z_{\ell_n}$. These points will
have to be fitted to
\begin{equation}
  f(z; z_{0_\ell},\alpha_\ell,\gamma_\ell) = \frac{1}{\alpha_\ell}(z-z_{0_\ell})^2 +
  \frac{\gamma_\ell}{\alpha_\ell}
\label{fit.0}  
\end{equation}
to obtain $z_{0_\ell}$, $\alpha_\ell$ and $\gamma_\ell$. We notice
that although $f$ is a quadratic,  we can use linear regression techniques
after a simple substitution. Following \cite{King} and his algorithm
to fit a hyperbola, we will assume that the focus position lies somewhere
in between $z_{\ell_1}$ to $z_{\ell_n}$. Thus, the set of all possible
focuser positions is simply
${\cal Z}_{0_{\ell}} = \left\{ z_{\ell_1}, z_{\ell_1}+1, z_{\ell_1}+2,
  \ldots, z_{\ell_n} = z_{\ell_1} + P \right\}$. We can then define a
new variable $\zeta_\ell(z;j) = [z-(z_{\ell_1}+j)]^2$ and transform
$f$ into a linear function in $\zeta_\ell$
\begin{equation}
    f(\zeta_\ell(z;j);\alpha_\ell(j),\gamma_\ell(j)) = \frac{1}{\alpha_\ell(j)}\zeta_\ell(z;j) +
  \frac{\gamma_\ell(j)}{\alpha_\ell(j)}
\label{fit.0a}  
\end{equation}
Now, we can apply simple linear regression (see \cite{Seltman} for a
description of this method) to the data points in ${\cal A}_\ell$ for
each focuser position $(z_{\ell_1}+j)$ to obtain $\alpha_\ell(j)$ and
$\gamma_\ell(j)$.

After we do this, we have $P+1$ solutions
$\{(z_{\ell_1}+j, \alpha_\ell(j), \gamma_\ell(j))|\; j=0,\ldots,
P\}$ from the fit. We select the best focus position from this set by
calculating the reduced\footnote{Although we call
  $\chi^2_{{\cal A}_\ell}$ in Eq.~\ref{fit.1} the reduced $\chi^2$, it is
  not the usual definition. For the usual definition, see
  \cite{Bevington}.}  $\chi^2$
\begin{equation}
  \chi_{{\cal A}_{\ell}}^2(j)=\frac{1}{n}\sum_{k=1}^{n}
\left[ \frac{1}{f(\zeta_\ell(z_{\ell_k};j);\alpha_\ell(j),\gamma_\ell(j))}
  -\frac{1}{Y_{\ell_{k}}}\right]^2
  \label{fit.1}  
\end{equation}  
for each $j=0,\ldots, P$. Let us suppose  $(z'_{0_\ell}, \alpha'_\ell,
\gamma'_\ell)$ gives us the the smallest reduced
$\chi^2$ for point set ${\cal A}_\ell$, then from here we can start
adding data points outside of ${\cal A}_\ell$ into it by using robust
estimators.
We do this by calculating the residual
\begin{equation}
  \epsilon_{\ell_{i}}=\frac{1}{f(z_i; z'_{0_\ell}, \alpha'_\ell,
\gamma'_\ell) }- 
 \frac{1}{Y_{i}}\label{eq:residuals}
\end{equation}
for each point $\left(z_i, Y_i\right)$ in
$\cal M$. From all the values $\{\epsilon_{\ell_i}\}$, a robust estimator, such
as the robust S-estimator $S(\{\epsilon_{\ell_i}\})$ from Rousseeuw, is
then calculated (see \cite{Sestimator} for a detailed description of
this estimator). For each data point $U=\left( z_u,Y_u\right)\in\cal M$, for
$u= 1,\ldots,N_{\cal M}$  so that  $U \not\in {\cal A}_\ell $, we can compute the
quantity
\begin{equation}
   T_U=\frac{|\epsilon_{\ell_q}-{\rm
      median}(\{\epsilon_{\ell_i}\})|}{S(\{\epsilon_{\ell_i}\}),\label{breakdown}}
\end{equation}
where
${\rm median}(\{\epsilon_{\ell_i}\})$ is the median of all the values
$\{\epsilon_{\ell_i}\}$. If $T_U$ is smaller than some user defined
tolerance (usually $2.5$ -- $3$), the point $U$ is added to the set
${\cal A}_\ell$. Thus, the size of ${\cal A}_\ell$ can increase, i.e.~$n\rightarrow N_{{\cal A}_\ell}$.

At this point, we have completed our work with the point set
${\cal A}_\ell$. We restart this procedure with another selection of
$n$ points from $\cal M$. The above procedure is repeated until we
have exhausted all $N_{\cal M} \choose n$ combinations. We are then
left with sets ${\cal A}_\ell$, for
$\ell=1,\ldots,{N_{\cal M} \choose n}$, each having
$N_{{\cal A}_{\ell}}\ge n$ points
$\left(z_{\ell_j}, Y_{\ell_j} \right)$ for
$j=1,\ldots,N_{{\cal A}_\ell}$, because additional inliers may have
been added to the initial combinations.

In order to increase the speed of the algorithm, we remove point
sets ${\cal A}_\ell$ that are duplicates.  
We do this by
associating a hash-value to each set ${\cal A}_\ell$. Then, we can generate a
hash table where we can look them up with a fast process to check whether
there are equal hash-values, i.e.~duplicate ${\cal A}_\ell$'s. If found, 
the duplicates are marked and ignored. In order to save memory, the
hash table may have to be cleared from time to time.

Finally, we perform simple linear regressions on each of the
remaining point sets, ${\cal A}_\ell$, for
$\ell=1,\ldots,J \le {N_{\cal M}\choose n}$, after duplicates have been
removed, by using Eq.~\ref{fit.0a} again.  The reduced $\chi^2$ error is
calculated for each ${\cal A}_\ell$ using Eq.~\ref{fit.1} but with
$n\rightarrow N_{{\cal A}_\ell}$.  At the end of the process, the algorithm
returns the ``found'' focus position, $z_0$, to be the value
that has the smallest $\chi_{{\cal A}_\ell}^2$ from the point sets
${\cal A}_1,\ldots,{\cal A}_J$.

We can increase the speed of the algorithm tremendously by using
parallel processing techniques (for example with OpenMP or by
using the new parallelism features in the C\texttt{++}17 standard).  We notice
that this algorithm has its worst performance when $n=N_{\cal M}/2$, because
$N_{\cal M} \choose n$ is at its maximum here which leads to the largest
possible number of attempted fits for a given data set.  For example,
a personal computer equipped with a Ryzen 9 3900X 12 core processor
can robustly fit models with $N_{\cal M}=24$ data points and
$n=N_{\cal M}/2=12$ using this algorithm in $3.17$~s. We can increase the speed
of the procedure further by using faster algorithms for the
S-estimator by \cite{fastSestimator}.

\subsection{Introduction of a shaping term}
\label{sec::shaping}
If we want to increase the precision of the fit, we can modify the
above algorithm by introducing a shaping term. This shaping term can
be thought of as a
heuristic of term~2 in Eq.~\ref{eq:doublesum}.
After the first step of our algorithm has fitted the measurement
results to $\frac{\alpha}{(z-z_0)^2+\gamma}$ and removed points which
are outliers with respect to this function, we can use the values for
$\alpha$, $z_0$ and $\gamma$ as initial values for the LM algorithm that fits the inlier data to the function: 
\begin{multline}
    g(z;\alpha,\beta,\gamma,z_0,\Theta)=\frac{\alpha}{(z-z_0)^2+\gamma}\\+\frac{\alpha}{(z-z_0)^2+\gamma}{\rm
      e}^{-\frac{\Theta}{(z-z_0)^2+\gamma}}\label{eq:gfunction},
  \end{multline}
  with the initial values $\Theta'_\ell=10\gamma'_\ell$ (See
  Appendix~\ref{app::dense} for the choice of ``10'').  The LM
  algorithm then returns the parameters
  $(z''_{0_\ell}, \alpha''_\ell,\gamma''_\ell,\Theta''_\ell)$, where
  $z''_{0_\ell}$ is the estimate of the focus position.
  
  We can then use the same statistical estimators as before in order
  to test, whether the points that were computed to be outliers with
  respect to initial parabola fit are still outliers when we use the
  LM algorithm with the shaping term. In general, the shaping term is
  a small correction to $ \bar{\cal P}_{\rm Fm}$. We have found that
  when this correction is included in the fit, the resulting curve has
  a smaller fit error and a lower number of outliers.  This has been
  confirmed by our experiments.
  
  In section~\ref{sec::Fmformula}, we have subtracted the parameter
  $\beta$ from $\bar{\cal P}_{\rm Fm}$. $\beta$ was determined by
  analyzing the white noise of the images in Fourier space. While our
  experiments show that doing this is sufficient for determining the
  focus position within the CFZ, we can improve the fit with LM with
  small corrections to this value of $\beta$. We start the process 
  with the initial correction value $\delta\beta=0$. Our
  experiments show that this reduces the fit error and the number of
  outliers of the curve fit even more.
  
  For the practical implementation, we have used the improvements of
  the LM algorithm described in \cite{Levenberg2}.
  
  \subsection{Large data sets}
  
   The algorithm described up to this point solves the trimmed least
  squares problem. Our experience running this algorithm on a personal
  computer shows that it becomes computationally difficult and time
  consuming once ${N_{\cal M} \choose n } > 2,704,156 $. For large
  data sets, we therefore choose a different method. In contrast to
  small data sets, we now start with a single fit of the data to
  $Y(\zeta,\alpha,\gamma)$. This means we use a
  single point set $\mathcal{A}$ which contains all the points from
  $1,\ldots,N_{\mathcal{M}}$ and not just some subset of points with
  indices $1,\ldots,n$ where $n\leq N_{\mathcal{M}}$.  Furthermore,
  the formula for the reduced $\chi^2_{\mathcal{A}_l(j)}$ in
  Eq.~\ref{fit.1} is modified.  One option for
  $\chi^2_{\mathcal{A}(j)}$ is to only compute from those points $k$
  where
\begin{equation}
\frac{1}{f(\zeta{(z_{_k};j)};\alpha(j),\gamma(j))}
  -\frac{1}{Y_{{k}}}
\end{equation} 
is not an outlier according to one of the statistical estimators
described in the sections above and the factor $1/n$ in front of
Eq.~\ref{fit.1} is given by the number of inliers. Another option is
that for the outliers, the contributions to
$\chi^2_{\mathcal{A}(j)}$  are given by
\begin{equation}
\left|\frac{1}{f(\zeta(z_{k};j);\alpha(j),\gamma(j))}
  -\frac{1}{Y_{{k}}}\right|
\end{equation}
whereas the errors of the inliers contribute quadratically to
$\chi^2_{\mathcal{A}(j)}$.

Obviously, the algorithm is different from the solution of the trimmed
least squares problem.  The values $\alpha(j)$ and $\gamma(j)$ are now
computed from a fit of all the points and the outliers only have a small
influence on the result for the estimated focus
position. Therefore, large errors that can be introduced by outliers
do not lead to a wrong estimate of the minimum of the parabola
anymore. Our tests with the 50 point samples (discussed in
section~\ref{sec::compareHFD}) show that this approach is sufficient
for large sets of measurement data. Since the computation still
depends in some way on all points that were measured, a second stage
fit with the LM that includes the shape term and small corrections of
$\beta$ is especially beneficial to this method.

\subsection{Comparison of fitting ease between Fm and HFDm}
\label{sec::compareFmHFDm}

In order to compare HFDm to Fm, we have adapted the application from
\cite{Benni} to provide fits and outlier rejection for data that is
modeled by Eq.~\ref{eq::powerfit}. The application in \cite{Benni}
makes several fits where various estimators are used within the RANSAC
inspired algorithm. This is done by replacing the S-estimator in
Eq.~\ref{breakdown} appropriately. The various estimators used in
\cite{Benni} have different properties that can be described by an
influence function. The latter determines the rate at which an
estimate changes after the insertion of an outlier. The influence
function of a robust estimator should be bounded because only then
will large outliers not have an impact at all. More precisely, it can
be characterized by the following properties:
\begin{itemize}
\item Efficacy ${\rm e}*$: The loss of efficacy of the estimator with
  respect to asymptotic variance. A high efficacy yields good
  estimates when the distribution of the samples is unknown.
\item Breakdown point $\epsilon*$: This is the maximum percentage of
  points to have limited results on the estimate.
\item Gross error sensitivity $\gamma*$: This describes the worst
  influence of a small amount of contamination of fixed size.
\item Rejection point $\rho*$: The distance from a point to an
  estimate where the influence function vanishes. An infinite
  rejection point means that all points are considered, whereas a
  small rejection point means that large outliers do not contribute to
  the estimate.  For example, if we use the S-estimator, we can calculate 
  $T_U$ from Eq. (\ref{breakdown}) for a given data point. We will reject
  it if $T_U>\tau$ where $\tau\in[2.5,3]$ is some pre-defined
  tolerance. 
\end{itemize}

\begin{table}
  \centering
  \caption{Some properties of various robust estimators }
\label{tab:tab1}
	\begin{tabular}{l|lll}
          \hline
  Estimator & \hfil$\epsilon*$\hfil & \hfil$\gamma*$\hfil & \hfil${\rm e}*$\hfil \\          
          \hline
Median absolute deviation (MAD) & 50\%        & 1.167                  & 37\% \\ 
S Estimator                     & 50\%        & 1.625                  & 58\% \\ 
Q Estimator                     & 50\%        & 2.069                  & 82\% \\ 
Biweight mid-variance           & 50\%        & \multicolumn{1}{c}{-}
                                                          & 86\% \\
\hline          
	\end{tabular}
      \end{table}

We have used our algorithm on several sets of measurement data. If the
HFDm is used, we noticed that due to the presence of many outliers, a
na\"ive linear fit of all points would often lead to a completely
wrong estimation for the focus position. This situation is greatly
improved by the RANSAC inspired algorithm presented above.

There are other systematic errors in HFDm that we have to consider as
well. HFDm has to first detect individual stars and determine the area
of their HFD. Unfortunately, the HFD often changes rapidly at the
focal point due to seeing. The necessary determination of the HFD may
also introduce systematic errors because it is difficult to measure
and compute the HFD correctly for small and dim stars from a briefly
exposed image. Such systematic errors that are introduced from the
image analysis are not entirely random and thus may not follow a
symmetric Gaussian distribution over the entire set of
images. Unfortunately, the stars also get less bright when de-focused
and their HFD is then even more difficult to compute. This makes the
HFD method mostly useful for images close to focus, but even here, the
HFD values are often disturbed by seeing. Furthermore, close to the
focal point, the HFD curve is empirically rather flat.

All the above problems inherent in HFDm make curve fitting often
difficult even if robust estimators in the RANSAC inspired algorithm
are used. For the S, Q, MAD and Biweight mid-variance estimators used
by the test application, the values for $\epsilon*$, $\gamma*$,
${\rm e}*$ can be found in \cite{huber2,
  Iglewicz,Hampel,huber1,wilcox,Sestimator,fastSestimator} and
summarized in Table~\ref{tab:tab1}. Because of their different
characteristics, we observe a rather broad spread of varying estimates
for the focus position when the RANSAC inspired algorithm is used with
different estimators on the HFDm data.

In contrast, the peak of $\bar{\cal P}_{\rm Fm}$ appears to be more
pronounced than the valley of the HFD curve. With the Fm data, the
outcome of the fit does not depend very much on $\gamma*$, because
small errors will always lead to a similar estimate for the focus
position. Furthermore, the steeper slopes of $\bar{\cal P}_{\rm Fm}$
increase the reduced $\chi^2$ of a fit that includes outliers
considerably. Therefore, the fit of the Fm data is less
sensitive to $\rho*$ of the estimator that is used
when compared to the HFDm.

Furthermore, Fm also filters out noise to some degree because it
contains one integration which acts like a low pass filter. And since
it does not depend on any star detection procedure, it has less
systematic non-Gaussian errors, which makes it less dependent on the
${\rm e}*$ of the estimator.

Unsurprisingly, we find that the estimators, each with their own
$\epsilon*$, $\gamma*$, ${\rm e}*$, and $\rho*$ values, when applied
to Fm data usually yields the same or very close values for the focus
positions.  In Table~\ref{tab:tab2}, we computed the standard
deviation of the focus positions found by the four estimators with
HFDm and Fm data. The small standard deviation of the Fm results when
compared to HFDm results shows that the robust fitting algorithm is
able to compute a more convincing estimate with the Fm data than the
HFDm data. This behavior has been observed in a number of our
measurements not shown in this paper. We will show one example in
section~\ref{sec::ngc7023}.

The fact that Fm data is usable with the four robust estimators
discussed here means that the fast but low efficacy estimator
(${\rm e}*$) MAD can be used in place of the slower S- or
Q-estimators.

Based on visual inspection of the data (humans are often better in
spotting outliers than machines), we note that despite the pronounced
peak at the focus position in Fm, the application of a simple linear
fit or Siegel's median slope would still often yield a wrong estimate
of the focus position due to the influence of the remaining
outliers. Thus it appears that despite the Fourier transformation, the
RANSAC inspired algorithm with a robust estimator is often essential
and cannot be entirely replaced by a simple linear fit or a median
slope fit.

\section{Examples}
\label{sec::examples}

We will test Fm with two refractor and a Schmidt Cassegrain (SCT)
telescopes first before testing it on the 2~m RCC.  The images
collected by these telescopes will be processed by both Fm and HFDm so
that we can compare them.  We will fit the data points generated by Fm
and HFDm (HFD's of all the images used in this paper are obtained from
an astrophoto\-graphy program called {\tt APT\/}) with the algorithm that
we have described in section~\ref{sec::fit}. The settings for all the
estimators are $T_U=3$ and the maximum number of outliers that can be
removed is 8. For the purposes of determining whether Fm agrees with
HFDm, we have chosen to use the S-estimator for all the examples below
rather than using all the estimators described in
Table~\ref{tab:tab1}. We will also calculate the critical focus zone
(CFZ) (\cite{CFZ}) for each telescope/camera combination. If the focus
positions found by Fm and HFDm are within the CFZ, we consider them to
be equivalent. Note: In these examples, we have chosen images that
contain nebul\ae, galaxies and globular clusters to show that Fm is
not affected by them.

\subsection{Refractor telescope I}
The first refractor used in the following tests is an Astro-Physics AP130GT
which has an aperture of 130 mm and focal ratio F/6.6 (measured). It
is equipped with with a MoonLite focuser ($0.269$~$\mu$m/step)
that has zero backlash. The imaging color CCD camera is a SBIG
STF8300C ($5.4\times 5.4$~$\mu{\rm m}^2$ square pixel). In these
measurements, the images have not been de-Bayered and each image is the
result of 8~s of exposure. The CFZ for this
setup is 108~$\mu$m for green light (510~nm). The CFZ in terms of step
size is 401.

\subsubsection{Refractor telescope pointed at a nebula (NGC7023)}
\label{sec::ngc7023}
      
\begin{figure*}
  \centering
    \includegraphics*[width=\textwidth]{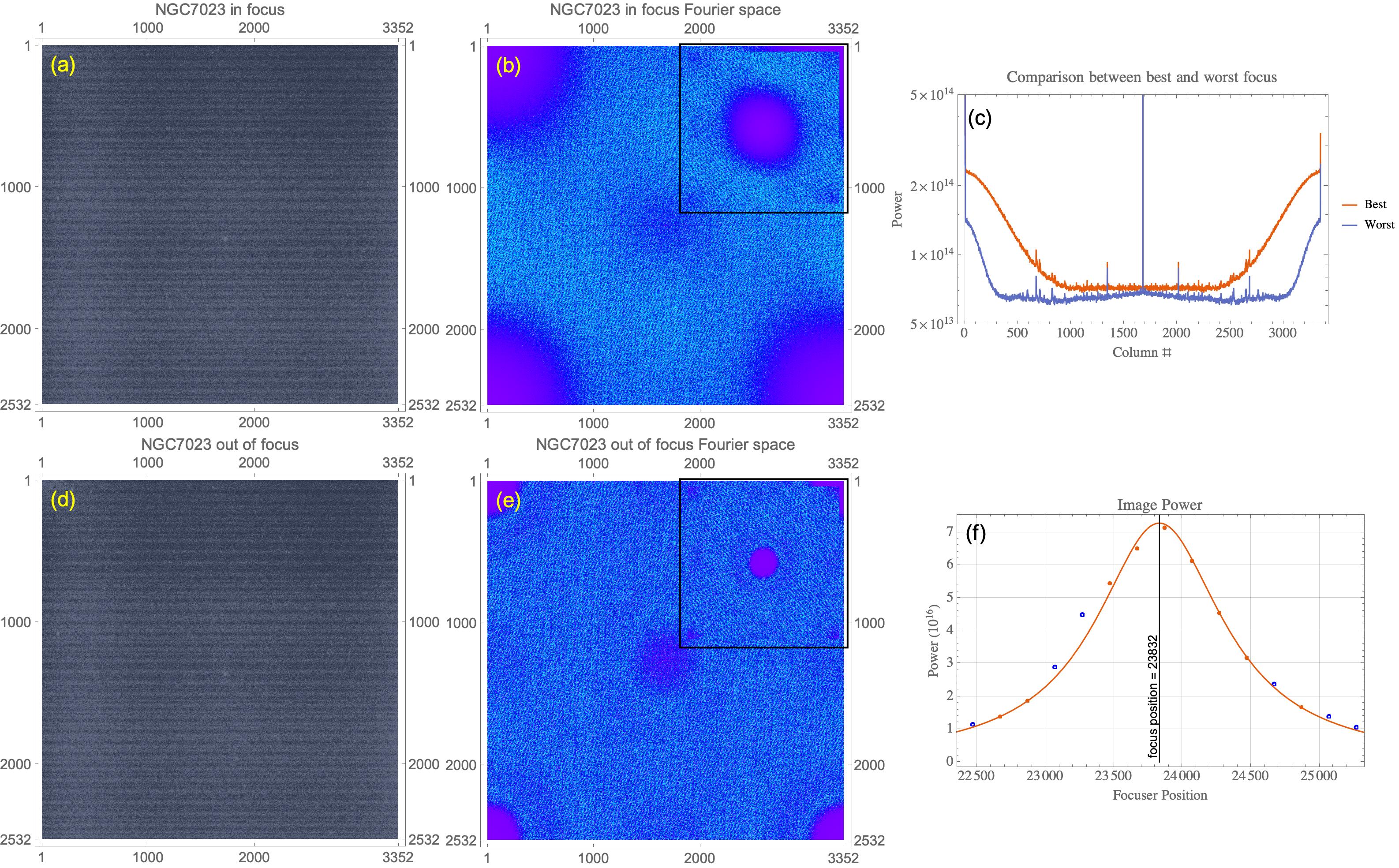}
    \caption{The images of NGC7023 (Iris Nebula) taken by our
      refractor I. The (a) in-focus and (d) out of focus images in
      real space are both lightly stretched. The nebula can actually
      be seen in our 8 s exposure. After applying a 2-D FFT to the
      real images, we obtain (b) and (e), which show that there are
      clear differences in the size of the lobes. The insets of (b)
      and (e) contain the same Fourier data except that zero frequency
      is at the center rather than at the corners. (c) The projection
      of the lobes into 1-D space are compared with a vertical
      logarithmic scale. (f) And when plotted as a function of focuser
      position and fitted to Eq.~\ref{fm.2} tells us where the focus
      position is.  The blue unfilled circles are the outliers
      identified by the fitting algorithm and ignored by the fit.}
  \label{gridngc7023.pdf}
\end{figure*}

We pointed our refractor I at the Iris Nebula (NGC7023) as a test of
Fm. The images in real and Fourier space are shown in
Fig.~\ref{gridngc7023.pdf}.\footnote{We have added an inset to all the
Fourier plots which has zero frequency at the center. This is 
for readers who are more comfortable with this type of
representation. We do not process the Fourier data in this shifted
zero format.}
The fit to Eq.~\ref{fm.2} for four types
of estimators are shown in Table~\ref{tab:tab2}. We can see that the
results for Fm are nearly identical and are within 5 focuser steps.
As a comparison, when we apply the HFDm data to the fit with each
estimator, the difference between each estimator
result can lie outside the CFZ. In fact, for the case of the
S-estimator, which we are using for the rest of the examples, the Fm
and HFDm results do not lie within the CFZ.  This example clearly
illustrates problems with the HFDm method that was discussed in
section~\ref{sec::compareFmHFDm}.

For completeness, we compare the focuser position results without
using outlier rejection in Table~\ref{tab:tab2a}. Again, the Fm
results have a much tighter spread than the HFDm results. When we look
at the last row of Table~\ref{tab:tab2} and \ref{tab:tab2a}, we see
that the Fm results are identical while it is not for HFDm.

\begin{table}
  \centering  
\caption{NGC7023 focusing results for Fm and HFDm with outlier rejection}
\label{tab:tab2}
\begin{tabular}{l|cccc}
  \hline\\
  Estimator & \hfil Fm\hfil & \hfil HFDm\hfil & \hfil Diff. \hfil  &\hfil Within\hfil\\
 & \hfil pos.\hfil&\hfil pos.\hfil &
                                     \hfil${\rm Fm} - {\rm HFDm}$\hfil  & \hfil CFZ?  \hfil\\
\hline
MAD                              & 23836    & 23928    & $-92$  & Y\\ 
S Estimator                     &23832    & 22952   & $880$  & N\\ 
Q Estimator                    &23836    & 23928    & $-92$  & Y\\ 
  Biweight     &23836   & 24072    & $-236$  & Y\\
  \hline
  Mean. and   & 23835 & 23720\\
  std. dev. of & 2 & 447 & &  \\
 results with est.  & & & &\\
\hline
\end{tabular}
\end{table}

\begin{table}
  \centering  
\caption{NGC7023 focusing results for Fm and HFDm without outlier rejection}
\label{tab:tab2a}
\begin{tabular}{l|cccc}
\hline 
Methods w/o & \hfil Fm\hfil & \hfil HFDm\hfil &
  \hfil Diff. \hfil  &\hfil Within\hfil
 \\outlier removal & \hfil pos.\hfil&\hfil pos.\hfil  & \hfil${\rm Fm} - {\rm HFDm}$\hfil  & \hfil CFZ?  \hfil\\
 \\
\hline
Lin. regression  & 23837    & 24043    & $-207$  & Y\\ 
  Siegel's median& 23832    & 24067 & $-235$&Y \\
   slope & & &\\
  \hline
  Mean and   & 23835 & 24055\\
  std. dev. of & 3 & 12 & &  \\
  results w/o est.  & & & &\\
  \hline
\end{tabular}
\end{table}

\subsection{Refractor telescope II}

The second refractor used in the following test is a Teleskop
quadruplet refractor 65/420 which has an aperture of 65~mm and focal
ratio F/$6.46$. It is equipped with a Sesto robotic focusing
motor connected to a rack and pinion focuser. The calibration of this
focuser is $0.7326$~$\mu$m/step.  The imaging CCD camera is a monochrome
ZWO ASI1600MM-P cooled camera ($3.8\times 3.8$~$\mu{\rm m}^2$
square pixel). In these measurements, we have used a red filter and
each image is the result of 8~s of exposure. The CFZ for this setup is
132~$\mu$m for red light (510~nm). The CFZ in terms of focuser
step size is 180.

\subsubsection{Refractor telescope pointed at M92 for a dense star
  field}
\label{sec::M92}

\begin{figure*}
  \centering
    \includegraphics*[width=\textwidth]{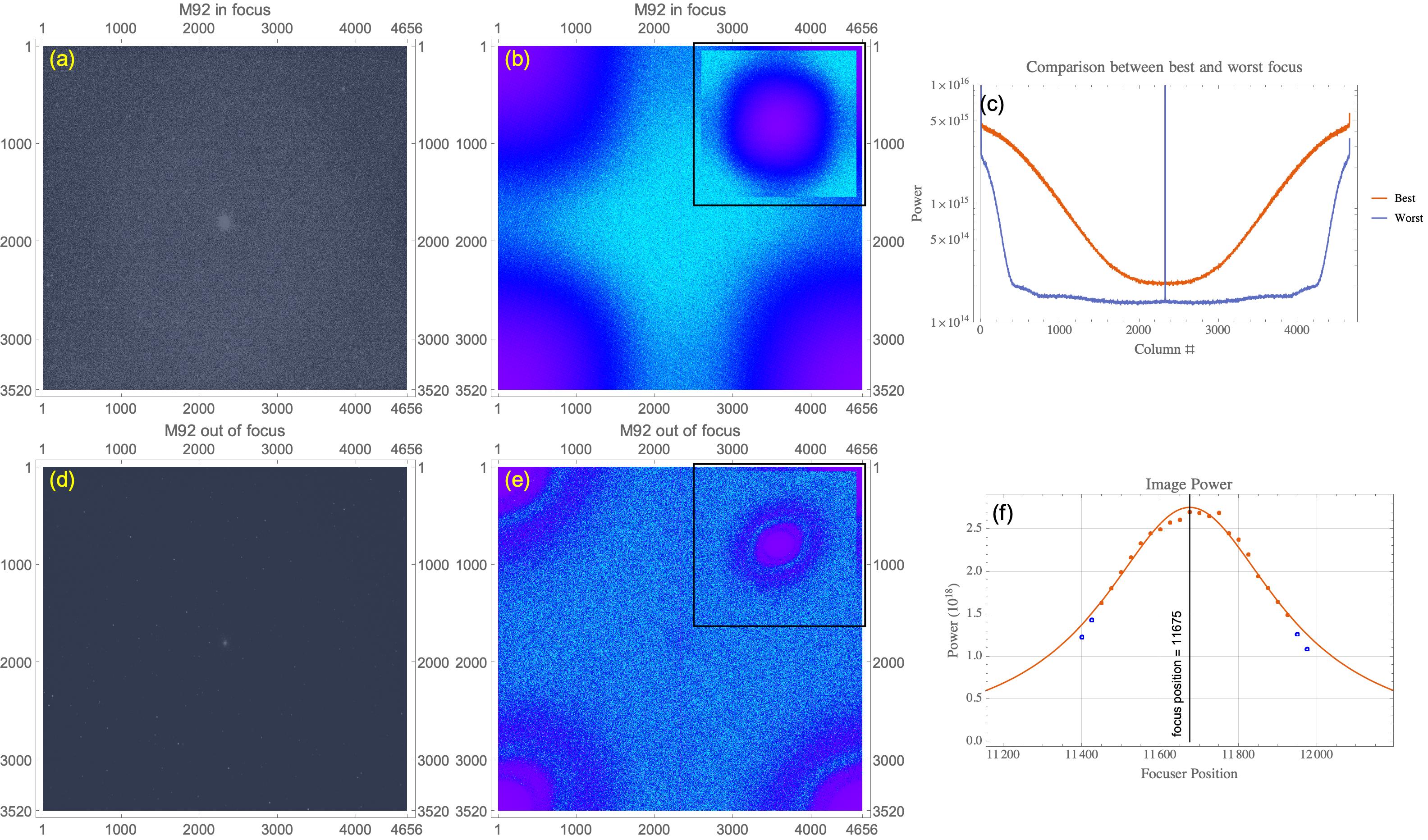}
     \caption{The images of M92 (globular cluster in Hercules) by our
      refractor II. In (f), we can see from the ``Image power'' plot that
      our fit finds the focuser position in a dense star field.}
  \label{gridM92.pdf}
\end{figure*}

We decided to test Fm for the case where there is a dense star field
because in this situation the assumption that the double sum in
Eq.~\ref{ph.3} vanishes does not apply. In fact, from our work in
section~\ref{sec::dense}, we need to check whether the power from a
dense star field fits well to the sparse star field approximation
given by Eq.~\ref{ph.5}.

We chose M92 (globular cluster in the constellation Hercules) for this
test. The results of our analysis is shown in
Fig.~\ref{gridM92.pdf}. The S-estimator finds only four outliers to
ignore and the fit function given by Eq.~\ref{ph.5} can still
find the focus position. The focus position with the S-estimator is
11675.  The fit to the HFDm data found the focus position to be
11693. The difference between the two methods is $18$~focuser
steps which is well inside the CFZ of 180 focuser steps. Thus,
both Fm and HFDm have found the same focus position with the
S-estimator.

The average of the focus positions found by the four estimators shown
in Table~\ref{tab:tab1} is $(11676\pm 1)$ for the Fm data, and
$(11692\pm 1)$ for the HFDm data. In this example, all four
estimators found results that are within CFZ for both the Fm and HFDm
data. And they have the same spread as well.

\subsection{Schmidt Cassegrain telescope}

\begin{figure*}
  \centering
    \includegraphics*[width=\textwidth]{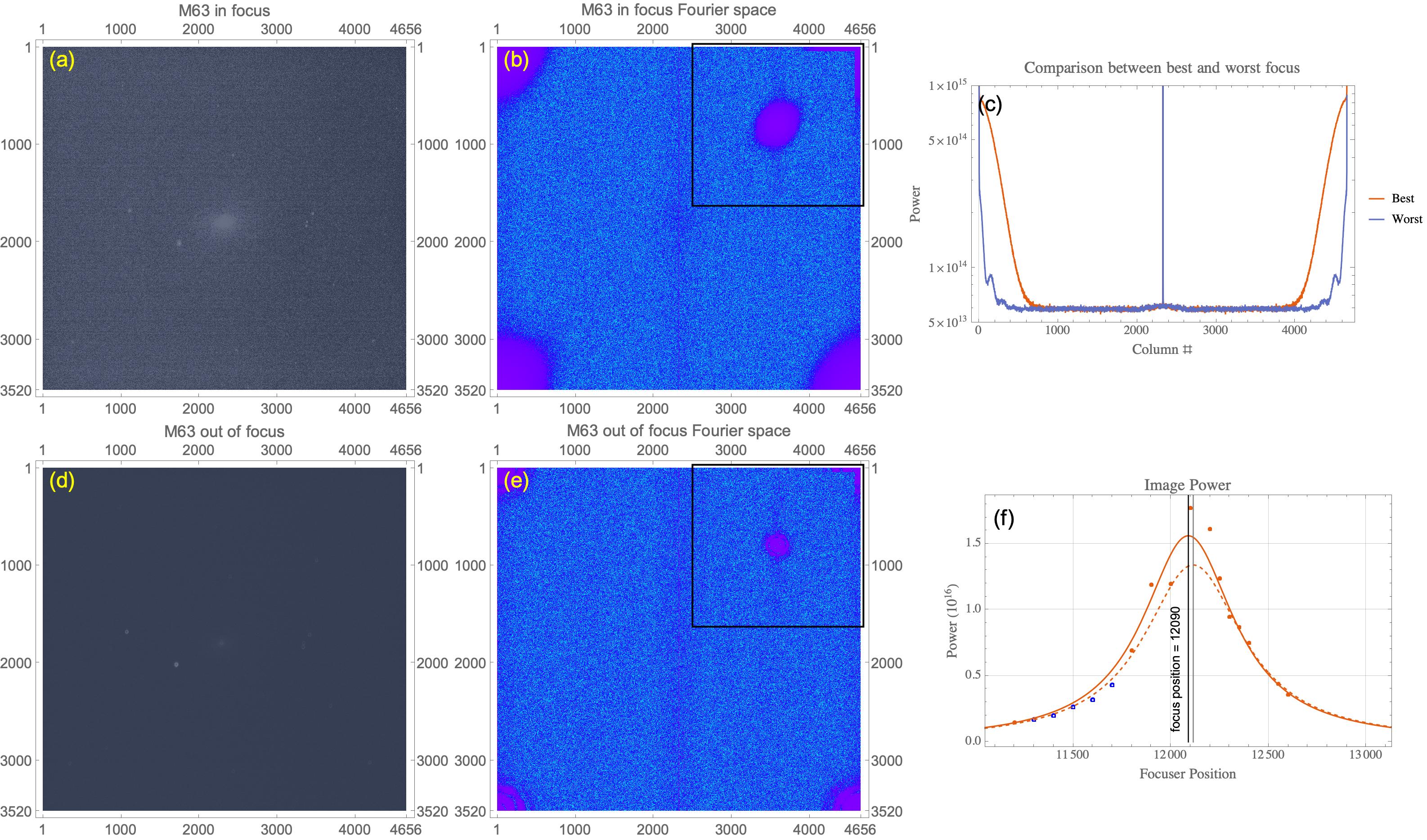}
    \caption{We pointed our SCT (Celestron EdgeHD 800) with a ZWO CMOS
      camera (ASI1600MM-P) at M63 (Sunflower Galaxy). The images shown
      here are after 7 s exposure.  The (a) in-focus and (d) out of
      focus images in real space are both lightly stretched. After
      applying a 2-D FFT to the real images, we see that there are
      clear bumps in the blue line of (c). These bumps come from the
      doughnuts.  In (f), the ``Image power'' plot contains two fits:
      a fit that ignores 5 outliers (continuous red line), and a fit
      that has no outliers (red dashed line).}
  \label{gridC8.pdf}
\end{figure*}

The SCT that we used in the following test is a Celestron EdgeHD 800
which has an aperture of 8{\tt "} and focal ratio
F/10. It is equipped with a DC motor focuser from JMI
. Since the JMI focuser does not use a stepper motor, the
steps reported here are emulated values. 
We do not have the calibration that tells us the travel distance of
the focuser per step because of the way the JMI focuser is connected
to the SCT focusing mechanism, and the idiosyncrasies of the SCT
focusing mechanism itself.  The imaging CMOS camera is a ZWO
ASI1600MM-P cooled camera ($3.8\times 3.8$~$\mu{\rm m}^2$ square
pixel).

The CFZ for this setup is 249~$\mu$m for green light (510~nm).
Unfortunately, since we do not have the focuser calibration, the best
we can do is to make an educated guess of the CFZ in terms of emulated
step size. We do this by noticing that any measurable change in the
HFD requires between $50 - 80$~steps. Therefore, we will use this
range as the estimate of the CFZ.

\subsubsection{SCT pointed at a galaxy (M63)}

We pointed our SCT at the Sunflower Galaxy (M63) for this test. A
feature of an SCT is that when it is de-focused, the star images look
like doughnuts because of the obstruction from the secondary
mirror. As expected, we can see from Fig.~\ref{gridC8.pdf}(d) that the
image of each star resembles a doughnut when de-focused. The doughnut
effect adds more bumps to the spectrum in Fourier space which we can
see in Fig.~\ref{gridC8.pdf}(c). We want to see whether Fm is affected
by this effect.

Unfortunately, the quality of the data is not as good when compared to
the refractor data. This is due to the entire focusing train of the
SCT not having proper linear motion that is reliably reproducible. But
despite these shortcomings, we can see from Fig.~\ref{gridC8.pdf}(f)
that the Fm data can be fitted to all the data points or with 5
outliers in the tail, i.e.~power from images that contain doughnuts,
removed. (Note: This is done in our program by specifying at most 8 or
0 outliers for the S-estimation algorithm.)  The focus position when
all the data points are used is 12115 emulated focus steps. While with
5 outliers ignored, we find the focus position to be at 12090 emulated
focal steps. The difference between the two results is 25 emulated
focus steps which is within our estimate of the CFZ for this
telescope/camera combination. This result means that Fm is not
affected by doughnuts in the tails.

For comparison, when we use the HFDm data, the fit finds the focus
position to be at 12091. Thus, both Fm and HFDm data have yielded
essentially the same focus position that is within our estimated CFZ.

The average of the focus positions found by the same estimators shown
in Table~\ref{tab:tab1} is $(12106\pm 9)$ for the Fm data, and
$(12091\pm 0)$ for the HFDm data. In this example, all four estimators
found results that are within the estimated CFZ, and for the HFDm
data, they found the same focus position with zero spread.

\subsection{2~m Ritchey-Chr\'etien-Coud\'e telescope}

\begin{figure*}
  \centering
    \includegraphics*[width=\textwidth]{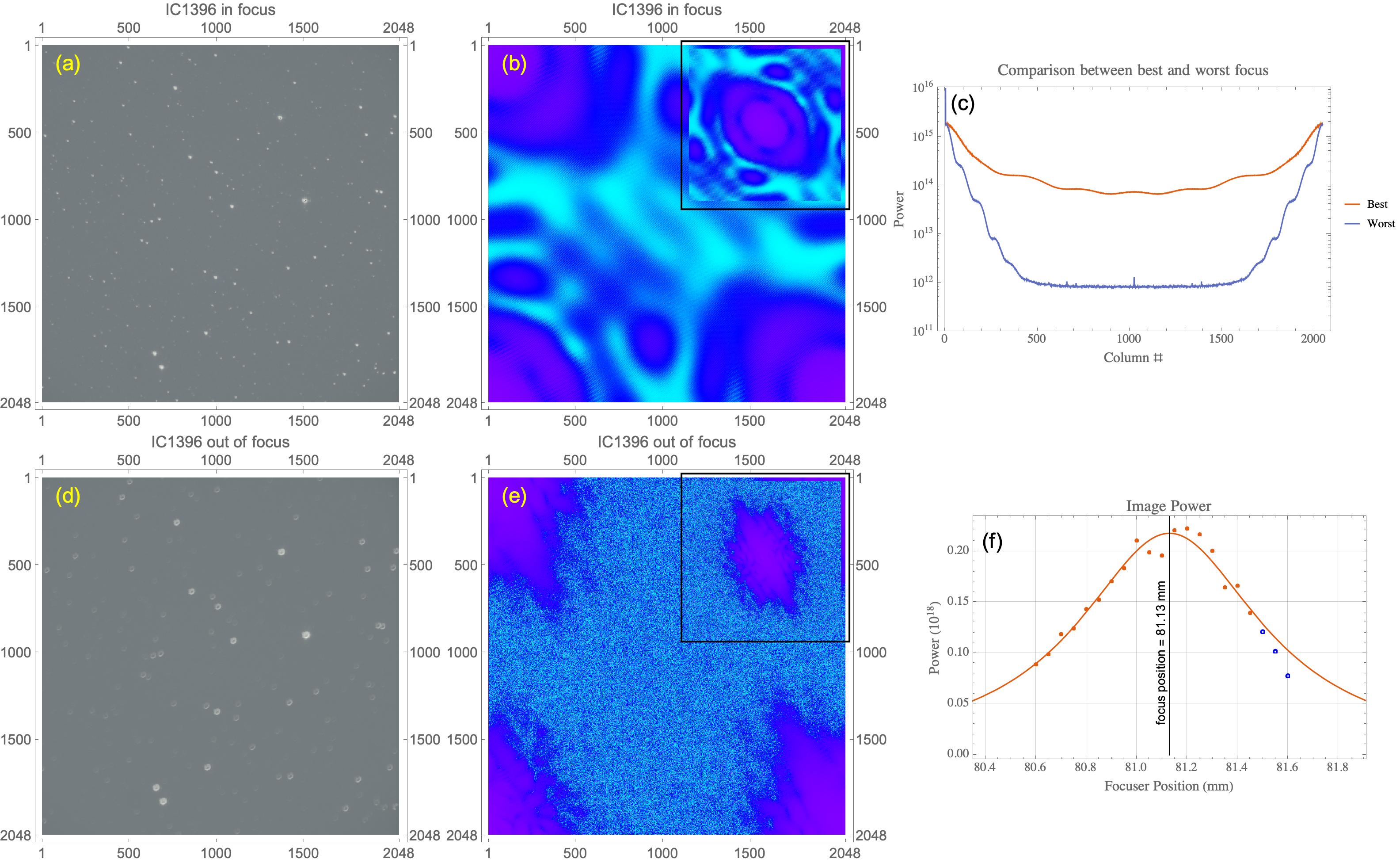}
    \caption{The 2~m RCC was pointed at IC1396 (Elephant Trunk
      Nebula). The images shown here are after 5 s exposure.  The (a)
      in-focus and (d) out of focus images in real space are not
      stretched here.  After applying a 2-D FFT to the real images, we
      see that the results are quite different from the those taken by
      small telescopes. In (f), the ``Image power''  fit has 3 outliers.}
  \label{gridIC1396.pdf}
\end{figure*}

After our successful Fm tests with small telescopes, we had the
opportunity to test Fm on the 2~m RCC at the Rozhen NAO.
R.~Bogdanovski (Rozhen NAO) and M.~Minev (Rozhen NAO) collected images
for us.  The RCC has a focal ratio F/8 and is equipped with a focuser
that has an absolute encoder. The absolute encoder has a resolution of
10~$\mu$m. The focuser has no backlash because of the absolute
encoder.  The CCD camera used for these images has a pixel size of
$(13.5\times 13.5)$~$\mu{\rm m}^2$ square pixel.  The CFZ for this
setup is 159~$\mu$m for green light (510~nm).

\subsubsection{RCC pointed at a nebula (IC1396)}

The RCC was pointed at the Elephant Trunk Nebula (IC1396) for this
test. Like the SCT, when it is de-focused, the star images look like
doughnuts because of the obstruction from the secondary mirror. The
collimation of the RCC was not perfect when the data was taken and so
the stars are not point like even when in focus. See
Fig.~\ref{gridIC1396.pdf}(a). There is also a lot of structure in the
Fourier~space data shown in Figures~\ref{gridIC1396.pdf}(b) and
\ref{gridIC1396.pdf}(e). It is interesting that
Fig.~\ref{gridIC1396.pdf}(c) shows that these structures are smoothed
out because of the first integral in Fm. However, the doughnut
structure of the stars are preserved because these curves clearly show
bumps. But these bumps do not affect the ``Image Power'' from our plot
shown in Fig.~\ref{gridIC1396.pdf}(f). The fit to the data with Fm
finds the focus position to be $81.13$~mm. When we apply HFDm to 
the same images, the focus position is $81.11$~mm. These two
results show that both Fm and HFDm agree because they are within the
CFZ.

The average of the focus positions found by the four estimators shown
in Table~\ref{tab:tab1} is $(81.130\pm 0.000)$~mm for the Fm data,
i.e.~all four estimators found the same results for Fm. Using HFDm,
the four estimators found the focus position to be
$(81.111\pm 0.004)$~mm. In this instance, both Fm and HFDm agree
because both are within the CFZ.

\section{Effects of Saturation}
\label{sec::saturated}

\begin{figure*}
  \centering
    \includegraphics*[width=\textwidth]{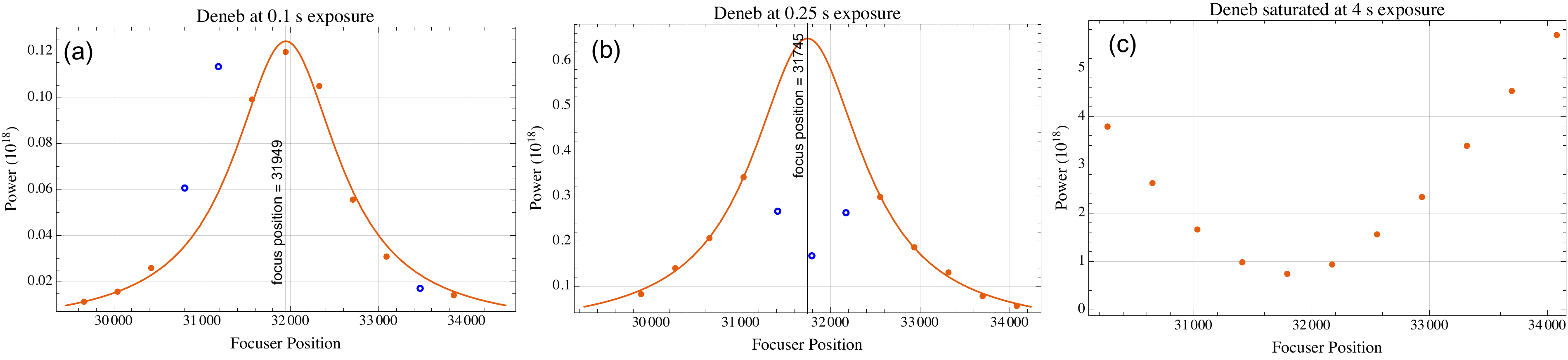}
    \caption{These plots show the effect of performing Fm with Deneb
      in the imaging frame. (a) For a short exposure of 100~ms, Fm
      finds the focusing position. (b) When the exposure time is
      increased to 250 ms, Deneb becomes saturated near focus which
      creates a dip in the Lorentzian. (c) Finally, when the exposure
      time is 4~s, the expected maximum at focus becomes a minimum.}
  \label{saturated.pdf}
\end{figure*}

Fm does not work well when there are saturated stars in the image. We
show an an example of the effect of a saturated star by pointing a
Takahashi FSQ106 refractor at Deneb (1.25 magnitude).  The FSQ106, has
an aperture of 106~mm and focal ratio F/5. It is equipped with with a
MoonLite focuser ($0.269$~$\mu$m/step) that has zero backlash. The
imaging color CCD camera is a SBIG STF8300C ($5.4\times
5.4$~$\mu{\rm m}^2$ square pixel).

The effect on Fm as the exposure time is increased is shown in
Fig.~\ref{saturated.pdf}. For a short exposure time of 100~ms, Fm
behaves as expected. But as the exposure time is increased to 250~ms, a dip
appears. This is because the image of Deneb is saturated near focus.
Finally, when Deneb is overexposed at 4~s, the expected
Lorentzian peak completely vanishes and is replaced by a valley.

The reason can be seen in Fig.~\ref{gaussian.pdf} where we have drawn
a dashed line to represent the saturation threshold. Let us suppose
that the star image is truncated above this threshold,
i.e.~saturation. Then, it is immediately obvious that its power
contribution is reduced because its peak is truncated. The
power is further reduced as the star becomes more in focus because the
base of the star image shrinks and most of the power contribution is
above the threshold level.

Therefore, during focusing, we have to be careful not to have any
saturated stars in the image or to apply an algorithm to damp down its effect. 

One way to do this is to reduce the
exposure time. But reducing the exposure to very short values has some
drawbacks. It reduces the contributions of the other stars to the
power function $\bar{\cal P}_{\rm image}$. Also, shorter exposures
make the process more prone to seeing fluctuations.  Therefore, we are
currently testing software based methods that prevent the creation of
the valley shown in Fig.~\ref{saturated.pdf}(c)  when saturated stars are in the
image.

One solution is to to define a new maximum which is smaller,
e.g. $\rm{max}_{\rm new}=\varphi \cdot \rm max+\rm min$, where $\rm max$ is the
largest value supported by the image format and $\rm min$ is the lowest value of any pixel in the given image. The variable $\varphi$ is some numerical constant that has been shown to deliver good results for the  bit depth of the used image format. 

We first check whether pixels with a value of $\rm max$ are in the
image. If that is the case, we search for all pixels whose value is
$\geq \rm{max}_{\rm new}$ and replace them by $\rm{max}_{\rm new}/2$,
which is the average of $\varphi\cdot\rm max$ and $\rm min$. For the
16 bit images where we tested this procedure, we found that a value of
$\varphi=0.01$ delivered good results. The method gives both a smooth
outer region of the saturated star and a flattened profile that is no
longer saturated. For our example with the saturated star Deneb, this
method yields a power curve whose fit has the same focus position that
would be computed from parts of the image without the saturated star.

\section{Focusing simulation}
\label{sec::compareHFD}

One way to compare HFDm and Fm is to simulate multiple focusing
runs by selecting a smaller subset of images from a much larger image data
set. We can do the following:
\begin{enumerate}
  \item We take a large series of images that are spaced less
    than the CFZ. We chose the starting and ending focuser positions
    so that they  bracket the focus
    position. Furthermore, these images are  taken as quickly as possible so that the ambient
    temperature does not change during this process.
  \item We define the best focus position to be the image that has the
    largest number of identifiable stars.  
  \item We choose a fixed focuser increment that will allow us to
    select a much smaller subset of images. These selected images are
    spaced consecutively at the focuser increment.
  \item We fit these images to both HFDm and Fm to see how well they
    match to the best focus position.
  \end{enumerate}

\begin{figure*}
  \centering
    \includegraphics*[width=\textwidth]{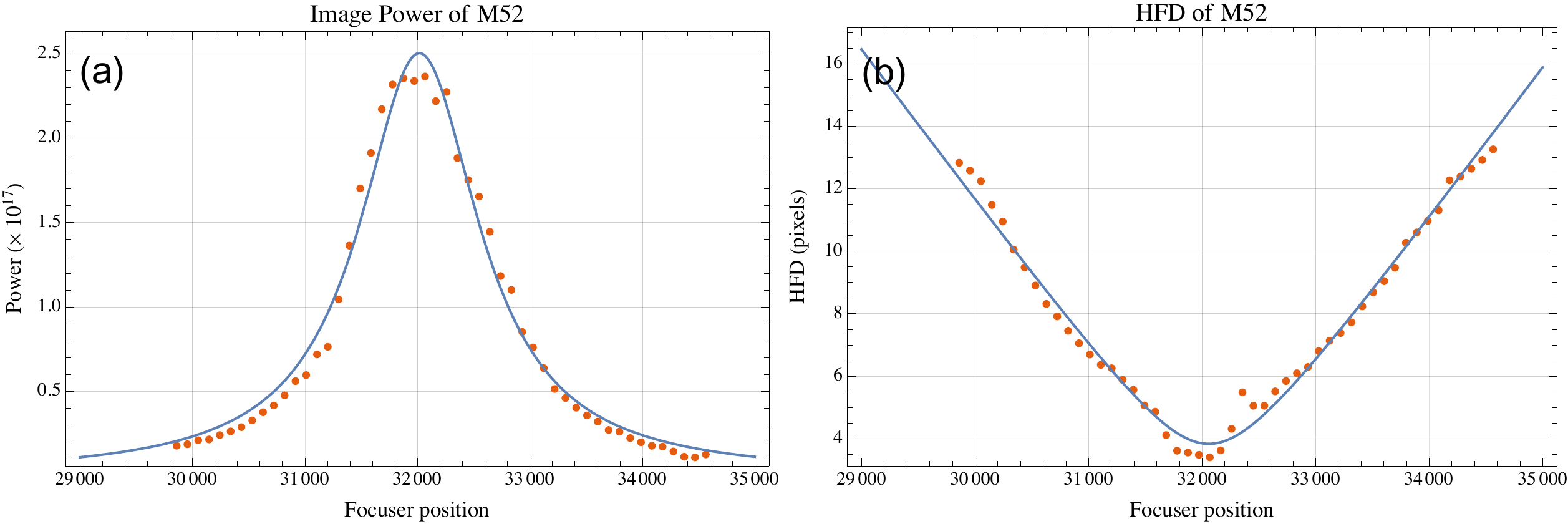}
    \caption{(a) shows the plot of 50 points of image power fitted to
      Eq.~\ref{cfd.1} and (b) shows the HFD fitted to
      Eq.~\ref{cfd.2}. No outliers were removed in these fits.}
  \label{g34.pdf}
\end{figure*}

The telescope that we will use for this experiment is the same Takahashi
FSQ106 that was described in section~\ref{sec::saturated}.  In these measurements,
the images have not been de-Bayered and each image is the result of
4~s of exposure. The CFZ for this setup is 62~$\mu$m for green light
(510~nm). The CFZ in terms of focuser step size is 230.

We pointed our telescope to a star cluster (M52) and took 50 images
starting at focuser position 34564 and ending at 29860 with a focuser
step size of $\Delta z = 96 \approx {\rm CFZ}/2$. Figure~\ref{g34.pdf}
shows the power and HFD of the images fitted to the following
equations:
\begin{align}
  {\cal{F}}_{\rm Fm} (z; \alpha, \gamma, z_0)&=
                                              \frac{\alpha}{(z-z_0)^2+\gamma}
\label{cfd.1}
  \\
  {\cal{F}}_{\rm HFDm}(z; \alpha, \beta,\gamma, z_0)&=
                                               \beta\sqrt{1+\frac{(z-z_0)^2}{\alpha^2}}
                                               +\gamma
\label{cfd.2}                                               
\end{align}
without the removal of any outliers. In the above equations,
Eq.~\ref{cfd.1} describes a Lorentzian and Eq.~\ref{cfd.2} describes a
hyperbola. The focus position found by Fm is 32018 and HFDm is 32060.

We used a plate solving program, {\tt ASTAP\/}, to determine the
number of stars in each image. The image with the largest number of
stars is when the focuser is at $z_{\rm max\; stars} = 31972$. We will
define $z_{\rm max\; stars}$ as the best focuser position. When we do
this, we can see that the difference between Fm and the best focuser
position is 46 steps and HFDm and the best focuser position is 88
steps. So, even although both Fm and HFDm found focus positions that
are within the CFZ of $z_{\rm max\; stars}$ position, we can see that
Fm is closer.

\subsubsection{Simulation}

Now, we can simulate a typical focusing run by selecting 5 to 8 images
per run. We constrain the selection of the images of each run by
making sure that each selection contains at least one image from the
first 10 images and the last image is within the last 10 images. We
constrained the selections this way so that each run brackets the best
focus position and not skewed to one side.

\begin{figure}
  \centering
    \includegraphics*[width=\columnwidth]{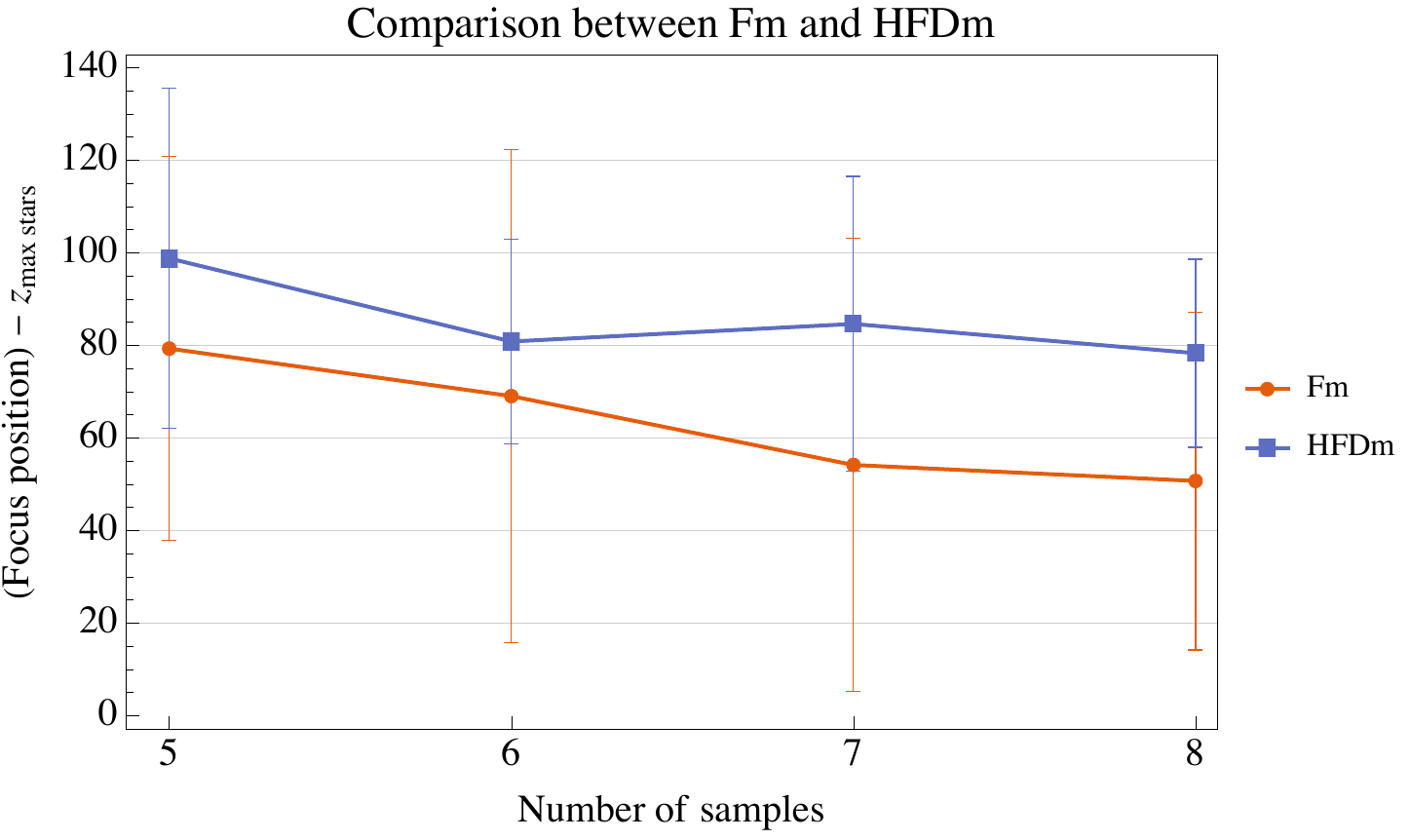}
    \caption{This figure shows the results of a simulated focusing run
      which samples 5 to 8 images from a population of 50 images. It
      is clear that the mean of the Fm focus positions are always
      closer than HFDm to the maximum star focus position.  But the difference
      between both methods and w.r.t.~$z_{\rm max\; stars}$ are
      within the CFZ.  }
  \label{compareFmHFDM.pdf}
\end{figure}

Fig.~\ref{compareFmHFDM.pdf} shows the 
results of our simulation. We can see that the mean of the Fm focus
positions is always closer to $z_{\rm max\; stars}$ than the HFDm
focus positions. But, we have to note that the difference between both
methods and w.r.t.~$z_{\rm max\; stars}$ is within the CFZ, so in
principle, both methods agree where the best focus position is.
Finally, it is also interesting, that both methods have found focus
positions that are always larger than $z_{\rm max\; stars}$.

From this simulation, we can say that for best results, we should
have at least 7 samples for either methods. But 5 samples is
still adequate for finding the best focus position.

\section{Strengths of F\lowercase{m}}
\label{sec::HFDm}

Fm does not suffer from the problems HFDm which are listed below. Our
studies in section~\ref{sec::examples} show that when the estimators
shown in Table~\ref{tab:tab1} are applied to the data points generated
by Fm, the results always have a spread less than the CFZ. But when
the same estimators are applied to the data points generated by HFDm
the spread can be larger than the CFZ. Our simulated focusing runs in
section~\ref{sec::compareHFD}, also show that both Fm and HFDm are
comparable in performance, with perhaps a slight edge to Fm. However,
as we have discussed in section~\ref{sec::saturated}, we have to be
careful about not saturating any of the stars during focusing.

We summarize here the weaknesses of HFDm:
\begin{enumerate}
\item \uline{Required detection of stars} Star detection is absolutely
  required in HFDm because the HFD of each detected star has to be
  calculated. Unfortunately, due to CCD/CMOS read noise, sky glow noise,
  noise from light pollution, and hot pixels the star detection algorithm has to
  be hand tuned to only detect stars and not other structures that
  look like stars. In contrast, Fm does not use star detection.
\item \uline{Noisy HFD measurements} The problems with HFD
  measurements are the following:
  \begin{enumerate}
  \item as focus gets close, the HFD actually becomes noisier as its
    value becomes smaller.  This is because when the star width is
    narrower, its fractional error becomes larger.
  \item far away from focus, the out of focus stars can be too dim to
    get a good HFD measurement or even be detected as stars. From our
    experience, even if HFDm detects these stars, the values of HFD in
    the tails are nearly always wrong.
 \end{enumerate}
 These problems manifest themselves when we apply our robust
 estimators to the same data and they yield a much wider swath of
 focus positions than Fm. We have already seen this in
 section~\ref{sec::ngc7023}.
\item\uline{HFDm step size} The focuser step size between each image
  has to be tuned for every telescope/camera combination in HFDm. It
  cannot be too small or else there will not be a sufficient change in
  the HFD between each position. If it is too large, star detection
  may fail because each star becomes too de-focused. Consequently HFDm
  will fail because star detection has failed. In comparison, the Fm
  step size can be chosen to be half the CFZ for each telescope/camera
  combination.
\end{enumerate}

Due to the superior results of Fm for obtaining the focus point
estimation, we have written a small console application called {\tt
  FM\/}. It expects the path to a folder with a small number of images
in fit format versus focuser position. The power of each image is
calculated for each focuser position which is then used to compute the
best focus position with a user selected estimator algorithm. The user
also selects the parameters used by the estimators. We have limited
the maximum number of images required by {\tt FM\/} to less than 24
images to keep the algorithm fast under all circumstances. We have
found that {\tt FM} needs about 8 images for a good precise fit,
but usually less than 17 images is sufficient. The source code of our
application was published under MIT license at \cite{Benni}. It comes
with extensive documentation and can be compiled on Windows, Mac OS X
and Linux.

\section{Conclusion}
We have demonstrated that Fm works well for finding the focus position
for both small and large telescopes. In the context of automation, it
is automatic because it does not require tuning of any parameters
except for the setting of the exposure time. This is in contrast to
HFDm which not only requires setting of the exposure time, it also
requires proper tuning of both star detection and focuser step sizes.
In our experiments, we have found that, unlike HFDm, Fm shows a strong
degree of independence from the choice of estimator. This means that
the user can be confident that the focus position found by Fm is
independent of the choice of the estimator.  We have published a
library that exports the algorithms and methods described in our paper
as a C\texttt{++} API, see \cite{Benni}. The API was kept as simple as
possible so that bindings to other languages, for example, Python, can
be easily generated with semi-automated tools. Finally, we provide a
small console application for Windows, Mac OS X and Linux under MIT
license 
that uses the library to compute the best focus position of a
telescope.  In conclusion, with Fm, we have increased by one the
number of automatic focusing methods available for professional and
amateur telescopes alike.

\section*{Acknowledgments}

We wish to thank the following people:
\begin{enumerate}
\item Ivaylo Stoynov, author of the Astro Photography Tool ({\tt APT\/})
  application, who kindly allowed us to discuss our thoughts about
  AutoFocusing on the forum pages that he hosts. He also introduced us
  to Dr.~R.~Bogdanovski.
\item Dr.~Rumen Bogdanovski (Institute of Astronomy and NAO, Bulgarian
  Academy of Sciences) and co-author of INDIGO who arranged for some
  telescope time with the 2~m RCC for the collection of data shown in
  Fig.~\ref{gridIC1396.pdf}.
\item Milen Minev (Institute of Astronomy and NAO, Bulgarian
  Academy of Sciences) who collected the data shown in
  Fig.~\ref{gridIC1396.pdf}.
\item Dr.~James Annis (Fermilab Cosmic Physics Center) who read our
  paper and encouraged us to publish these ideas.
\item Dr.~Adam Popowicz (Silesian University of Technology, Institute
  of Automatic Control) for his suggestion of a focusing simulation
  that we discussed in section~\ref{sec::compareHFD}.
\end{enumerate}

\section*{Data Availability}

The data used in this paper is available upon request from the
authors.



\bibliographystyle{mnras}
\bibliography{FM_v4} 




\appendix

\section{Dense star fields}
\label{app::dense}

When the star field is dense like in a star cluster, then the
assumption that the double sum in Eq.~\ref{ph.3} vanishes does not
apply. Suppose the mean star distances between neighboring stars is $\sim
\sqrt{m}\bar\sigma_0$, where $m\lesssim 10$. Then when we look at one of the terms in the
double sum 
\begin{equation}
  {\rm e}^{-\frac{(x_j-x_k)^2}{4\bar\sigma^2}} \rightarrow
  {\rm e}^{-\frac{m\bar\sigma_0^2}{4\bar\sigma^2}}
  \label{app.1}
\end{equation}
When in-focus, $\bar\sigma= \bar\sigma_0$, the above becomes
\begin{equation}
  \hbox{in-focus} =   {\rm e}^{-\frac{m}{4}}
  \label{app.2}
\end{equation}
And when not quite in-focus, $\bar\sigma= \bar\sigma_0 + \Delta\sigma$
with $\Delta\sigma/\sigma_0\ll 1$, we have
\begin{equation}
  \begin{aligned}
  \hbox{slightly out of focus} &=
  {\rm e}^{-\frac{m\bar\sigma_0^2}{4(\bar\sigma_0+\Delta\sigma)^2}}\\
&  \approx   {\rm e}^{-\frac{m}{4}(1-2\Delta\sigma/\bar\sigma_0)} = {\rm
    e}^{-\frac{m}{4}}{\rm e}^{\frac{m\Delta\sigma}{2\bar\sigma_0}}
  \end{aligned}
  \label{app.3}
\end{equation}
Here, we notice that ${\rm e}^{\frac{m\Delta\sigma}{2\bar\sigma_0}}>1$ because $\Delta\sigma
> 0$ and ${\rm e}^{-m/4} $ is not close enough to zero since $m\lesssim 10$. Thus, the double sum
\begin{equation}
  \begin{aligned}
&\left(\sum_{j=1}^N\sum_{k\ne j,
  k=1}^N  
{\rm e}^{-\frac{(y_j-y_k)^2}{4\bar\sigma^2}}{\rm e}^{-\frac{(x_j-x_k)^2}{4\bar\sigma^2}}\right)_{\hbox{in
  focus}} < \\
&\quad\left(\sum_{j=1}^N\sum_{k\ne j,
  k=1}^N  
{\rm e}^{-\frac{(y_j-y_k)^2}{4\bar\sigma^2}}{\rm e}^{-\frac{(x_j-x_k)^2}{4\bar\sigma^2}}\right)_{\hbox{slightly
  out of   focus}}
\end{aligned}
\label{app.4}
\end{equation}
Therefore, for the case of a dense star field, the double sum of an in-focus
image can be smaller than the power of a slightly out of focus image.


\bsp	
\label{lastpage}
\end{document}